\documentclass[fleqn,usenatbib,useAMS]{mnras}

\usepackage{newtxtext,newtxmath}

\usepackage[T1]{fontenc}

\DeclareRobustCommand{\VAN}[3]{#2}
\let\VANthebibliography\thebibliography
\def\thebibliography{\DeclareRobustCommand{\VAN}[3]{##3}\VANthebibliography}

\usepackage{graphicx}	
\usepackage{amsmath}	
\numberwithin{equation}{section}

\title[Shadow image of Sgr A*]{Probing the Shadow Image of the Sagittarius A* with Event Horizon Telescope}

\author[Saurabh et al.]{
Saurabh,$^{1}$\thanks{E-mail: sbhkmr1999@gmail.com}
Parth Bambhaniya,$^{2}$ \thanks{E-mail: grcollapse@gmail.com}
 and Pankaj S. Joshi, $^{2}$ \thanks{E-mail: psjcosmos@gmail.com}
\\
$^{1}$P. D. Patel Institute of Applied Sciences, Charusat University, Anand, GUJ 388421, India\\
$^{2}$ International Center for Cosmology, Charusat University, Anand, GUJ 388421, India\\
}

\date{Accepted XXX. Received YYY; in original form ZZZ}

\begin{document}
\label{firstpage}
\pagerange{\pageref{firstpage}--\pageref{lastpage}}
\maketitle

\begin{abstract}
Recent observations of the Milky-way galactic center at various frequencies suggest a supermassive compact object. Generally, that supermassive compact object is assumed to be a `Black Hole', having more than four million solar masses. In this work, we study the observational appearance at $230$ GHz and probe the nature of Sagittarius-A* (Sgr A*) as the naked singularity. Here, we consider the first type of Joshi-Malafarina-Narayan (JMN-1) and Janis-Newman-Winicour (JNW) naked singularity spacetimes which are anisotropic fluid solutions of the Einstein field equations. Motivated by radiatively inefficient accretion flows (RIAF), we use an analytical model for emission and absorption coefficients to solve the general relativistic radiative transfer equation. The resulting emission is then utilized to generate images to predict the nature of the Sgr A* with synthetic Very Long Baseline Interferometry (VLBI) images from current and future Event Horizon Telescope (EHT) arrays. Three different EHT array configurations are being used to simulate the models of naked singularities and a black hole. This may have little effect on the baseline, but it would increase the u-v plane gridding, making it feasible to capture a better-resolved image. Therefore, it is quite interesting and useful for the upcoming shadow image of the Sgr A* to predict whether it is a supermassive black hole or a naked singularity.
\end{abstract}


\newpage

\maketitle

\section{Introduction}

Why black holes are astonishing so far? As its distinct signature of black holes is their event horizon, a one-way membrane in spacetime where things can fall in but not escape away and even light cannot escape. Almost every galaxy is thought to have a supermassive black hole at its central core area \citep{2013ARA&A..51..511K}.   There is, however, no conclusive proof at this time. The Event Horizon Telescope (EHT) captured the first shadow image of a supermassive black hole at the center of the M87 galaxy in April of 2019 \citep{2019ApJ...875L...5E}. EHT is now being upgraded to the next-generation Event Horizon Telescope (ngEHT) to observe the galactic center of our Milky Way (Sgr A*) and other potential radio sources. The shadow image of the M87 galactic center observed by EHT does not confirm the existence of an event horizon and hence a supermassive black hole. Other compact objects can also cast similar shadows. The shadows cast by compact objects such as black holes, naked singularities, grava-stars, and wormholes have been extensively studied in \citep{2019PhRvD.100b4018G,2021arXiv211101752Y,2019PhRvD.100b4014A,2019PhRvD.100b4020V,2019PhRvD.100b4055G,2013PhRvD..88h3532D,2013PhRvD..87j3505D,2020PhRvD.102d4042D,2015PhRvD..92h4005A, 2015MNRAS.454.2423A, 2015PhRvD..91l4020O, 2019EPJC...79...44S, 2021arXiv210613175P, 2014PhRvD..90j4013S,2020MNRAS.497..521O, 2021arXiv210512057S,2019Sci...365..664D}. 

On the other hand, GRAVITY, SINFONI, UCLA Galactic center groups are continuously observing the stellar motions of the `S' stars around our Milky-way galactic center \citep{2019Sci...365..664D,2018A&A...615L..15G,2017PhRvL.118u1101H,2020A&A...636L...5G}. The astrometric and spectroscopic data for the S2 star is available and is being used to calculate its orbital precession and radial velocity respectively. The precession of S2-star orbits and the appearance of a shadow would provide information on the galactic center's causal structure. The naked singularity is differentiated from the black hole in several of our recent articles on orbital precession and shadows \citep{2019arXiv190908873J,2019IJMPD..2830024D,2021EPJC...81..205B,2021PhRvD.103h4005B}. However, to distinguish a naked singularity from a black hole from an observational approach, significant theoretical and observational analysis is required.

In \cite{2019MNRAS.482...52S}, the shadows of Joshi-Malafarina-Narayan (JMN) naked singularity spacetimes are investigated, and the results are compared to a Schwarzschild black hole shadow. The JMN-1 naked singularity can be formed as a result of gravitational collapse with zero radial and non-zero tangential pressures. Within the general theory of relativity, it could be obtained as quasi-equilibrium configuration of the collapsing fluid \citep{2011CQGra..28w5018J}. For the range of characteristic parameter $0<M_0<2/3$ (i.e. called Model-I in this paper), for which there is no photon sphere, and hence no shadow is determined in JMN-1 naked singularity, yet an intriguing, full moon image is formed. However, JMN-1 naked singularity can cast a similar shadow for $2/3<M_0<1$ (i.e. called Model-II in this paper), as the Schwarzschild black hole cast a shadow. 

Moreover, another model of naked singularity which was proposed in 1968 by Janis-Newman-Winicour (JNW) can cast a nearly similar shadow as the Schwarzschild black hole does \citep{1968PhRvL..20..878J,2020EPJC...80.1017G,2019PhRvD.100b4055G,2020PhRvD.102f4027S}. The JNW naked singularity is a minimally coupled massless scalar field solution of the Einstein field equations. In \cite{2020PhRvD.102f4027S}, the Authors showed that as the scalar field charge $'q'$ increases, and thus the radius of the photon sphere $r_{ph}$ increases while the shadow radius decreases. Recently, in \cite{2021arXiv210914937S}, the authors proposed the rotating version of the JNW naked singularity, while satisfying all the energy conditions. These models can thus act as good alternatives and can be used to test/probe the nature of the supermassive compact object at the galactic center. 

It was generally believed that the shadows arise due to the presence of a photon sphere. Recently, in \cite{2020PhRvD.102b4022J}, the authors have introduced a new spherically symmetric naked singularity solution of the Einstein field equation that lacks a photon sphere nevertheless casts a shadow. The general criteria for a shadow to occur in the absence of a photon sphere are then determined for null-like and timelike naked singularities, where both types of singularities satisfy all the energy conditions \citep{2021PhRvD.103b4015D}. It has been determined that the event horizon and photon sphere are not essential for the formation of a shadow. The presence of the upper bound of null-geodesics' effective potential induces the formation of a shadow, according to these investigations \citep{2021PhRvD.103b4015D}. As a result, if the effective potential of null-geodesics of a spacetime has an upper bound, the spacetime can cast a shadow. 

Although the EHT observations of the 2017 campaign of M87 conclusively proved the presence of shadow-like structure, it remains to be a question what the central object really is. As the EHT will be succeeded by its advanced version, ngEHT \citep{2021ApJS..253....5R}, more existing and new sites are being added to the current version. For the 2021 campaign, 3 additional sites were added: Greenland Telescope (GLT), Kit Peak (KP) 12 meter, and Northern Extended Millimeter Array (NOEMA). Proposed for EHT 2025 observations, there would be an addition of 12 telescopes (see table -\ref{table:sites}) making a total of 22 sites observing simultaneously. This will not cause much effect on the baseline but it will increase the u-v plane gridding, filling up the Fourier domain and making it plausible to capture a more clear image. In addition to the ground stations, space-based sites can prove to be very helpful in making a much more resolved image and test these alternative models \citep{2021A&A...650A..56R}.

The highly dynamic nature of the Sgr A* makes it very difficult to capture a static image. Hence, it also makes difficult to model the nature of the accretion environment robustly. At the moment, the preferred explanation for low-luminosity active galactic nuclei and black hole X-ray binaries, including Sgr A* is Radiative Inefficient Accretion Flows (RIAF) at low accretion rates \citep{1995Natur.374..623N,2009ApJ...699..626H,2003ApJ...598..301Y}. Here, we will investigate the resulting images of JMN-1 and JNW naked singularities and the Schwarzschild black hole surrounded by accretion flows. Motivated by RIAF, we consider an analytic model for emission and absorption coefficients to solve the general relativistic radiative transfer equation. The resulting emission is then used to generate images to probe the nature of the Sgr A* with VLBI images from current and future EHT arrays using synthetic observations.
The purpose of this paper is to predict the images of Sgr A* using the prescribed models with three observational campaigns.

The paper is organized as follows: In section (\ref{singularity}), we introduced the JMN-1 and JNW naked singularities. We describe the radiative transfer model in section (\ref{sec:level1}). In section (\ref{sec:rtf}), we describe the ray-tracing formalism and the initial conditions we use to generate accretion images. In section (\ref{VLBI}), we present and discuss the results obtained from the synthetic VLBI observation of the images and images reconstructed with current and future EHT arrays configurations. Finally, the discussion and conclusion are given in section (\ref{result}).

\section{JMN-1 and JNW Naked Singularities}
\label{singularity}
Here, we will briefly review the solutions that have been proposed in the literature for different naked singularities, for e.g., the JMN-1 and JNW naked singularities. The spacetime metric for a static and spherically symmetric object in the generalised form can be written as,
\begin{align}
ds^2 = -g_{tt}(r)dt^2 + g_{rr}(r)dr^2 + g_{\theta\theta}(r)d\theta^2 + g_{\phi\phi}(r)d\phi^2,
\end{align}
where, $g_{\mu\nu}$ are the metric tensor components and these are function of $r$ only.
\subsection{JMN-1 naked singularity spacetime}
The JMN-1 naked singularity can be formed as an end state of gravitational collapse with zero radial pressure and non-zero tangential pressures. This equilibrium configurations of the collapsing fluid within the general theory of relativity are extensively studied in \cite{2011CQGra..28w5018J}.  The JMN-1 naked singularity spacetime is described by the following metric tensor components, 
\begin{align}
   g_{tt}^{jmn} &= (1-M_0)\left(\frac{r}{R_b}\right)^{M_0/(1-M_0)},\\
  g_{rr}^{jmn} &= 1/(1-M_0),\\
   g_{\theta\theta}^{jmn} &= r^2,\\
   g_{\phi\phi}^{jmn} &= r^2\sin^2{\theta},
   \label{JMN1metric} 
\end{align}
where, $M_0$ and $R_b$ are positive constants. Here, $R_b$ represents the radius of the distributed matter around the central singularity and $M_0$ should be within the range $0<M_0<1$. The JMN-1 spacetime contains a curvature singularity at $r = 0$. The stress-energy tensor of the JMN-1 spacetime gives the energy density $\rho$ and pressures $p$ as,
\begin{equation}
    \rho=\frac{M_0}{r^2}, \; \; \;
    p_r=0, \; \; \;
    p_{\theta}=\frac{M_0}{4(1-M_0)}\rho.
\end{equation}
Note that in JMN-1, the collapsing fluid is supported only by tangential pressure, and it can be verified that all the energy conditions are satisfied by this spacetime \cite{2011CQGra..28w5018J}. 

The spacetime metric is modeled by considering a high-density compact region in a vacuum, which means that the spacetime configuration should be asymptotically flat. Therefore, if any spacetime metric is not asymptotically flat, we must match that interior spacetime to asymptotically flat exterior spacetime with a particular radius. The JMN-1 naked singularity spacetime is not asymptotically flat. Hence we can smoothly match this interior spacetime to exterior Schwarzschild spacetime at $r=R_b$ as,
\begin{equation}
    ds^2 = -\left(1-\frac{M_0 R_b}{r}\right) dt^2+\left(1-\frac{M_0 R_b}{r}\right)^{-1}dr^2+r^2d\Omega^2,
    \label{matchSCH}
\end{equation}
where, $\text{d}\Omega^2 = d\theta^2+\sin^2\theta d\phi^2$ and $M=\frac{1}{2}M_0 R_b$ is the total mass of the compact object. The extrinsic curvatures of JMN-1 and Schwarzschild spacetimes are automatically smoothly matched at $r=R_b$, since the JMN-1 spacetime has zero radial pressure \cite{2019PhRvD.100l4020B}. Here, the JMN-1 naked singularity is considered as Model-I in which we have defined the range of characteristic parameter $M_0$ within $0<M_0<2/3$ and for $2/3<M_0<1$, it is considered as Model-II. Note that, these ranges are determined theoretically based on the presence of a photon sphere.

\subsection{JNW naked singularity spacetime}
In 1968, Janis, Newman and Winicour obtained a minimally coupled mass-less scalar field solution of the Einstein field equations \citep{1968PhRvL..20..878J}, and independently by Wyman \citep{1997IJMPA..12.4831V}. The Lagrangian density of minimally coupled scalar field is given as,
\begin{equation}
\mathcal{L}=\sqrt{-g}\left(\frac12\partial^{\mu}\Phi\partial_{\mu}\Phi-V(\Phi)\right)\,\, ,
\end{equation}
where $\Phi$ is the scalar field, $V(\Phi)$ is the scalar field potential and $\sqrt{-g}$ is the determinant of metric. The minimal coupling conditions are defined as, 
\begin{equation}
    R_{\mu\nu}-\frac12 R g_{\mu\nu}=\kappa T_{\mu\nu}, 
   \end{equation}
\begin{equation}
    \Box\Phi(r)=V^{\prime}(\Phi(r)), 
\end{equation}
where $R$ is the Ricci scalar, $R_{\mu\nu}$ is the Ricci tensor, $T_{\mu\nu}$ is the energy-momentum tensor, and $g_{\mu\nu}$ is the metric tensor components. $\kappa$ is a constant parameter, defined as $8\pi G/c^4$. The Energy-momentum tensor ($T_{\mu\nu}$) for minimally coupled scalar field can be written as,
\begin{equation}
    T_{\mu\nu}=\partial_{\mu}\Phi\partial_{\nu}\Phi-g_{\mu\nu}\mathcal{L}.
\end{equation} 
One of the solution of this mass-less scalar field background is the Janis-Newman-Winicour spacetime which is the subject of interest in this work. The metric tensor components of the JNW spacetime are given as,
\begin{align}
   g_{tt}^{jnw} &= -\left(1-\frac{b}{r}\right)^\gamma,\\
  g_{rr}^{jnw} &= \left(1-\frac{b}{r}\right)^{-\gamma},\\
   g_{\theta\theta}^{jnw} &= r^2\left(1-\frac{b}{r}\right)^{1-\gamma},\\
   g_{\phi\phi}^{jnw} &= r^2\left(1-\frac{b}{r}\right)^{1-\gamma}\sin^2{\theta},
   \label{JNW1metric} 
\end{align}
where, $b=2\sqrt{M^2+q^2}$ and $ \gamma=\frac{2M}{b}$. Here, $M$ and $q$ represents Arnowitt-Deser-Misner (ADM) mass and scalar field charge respectively. As `$b$' is positive and greater than $2M$, we can write $0<\gamma<1$. The JNW spacetime contains a singularity at $r=b=2M/\gamma$ and it has a photon sphere for $\gamma > 1/2$, which is dubbed as `Model-III' for our current work. The mass-less scalar field can be written as,
\begin{equation}
    \Phi=\frac{q}{b\sqrt{4\pi}}ln\left(1-\frac{b}{r}\right)\,\, .
\end{equation}
The JNW spacetime is considered as an extension of the Schwarzschild spacetime when we include the minimally coupled massless scalar field and it smoothly translates to the Schwarzschild solution by considering the value of the scalar field charge $q=0$ and $\gamma=1$. Note that the JNW spacetime is asymptotically flat, therefore, we need not match this spacetime with the Schwarzschild spacetime.

\section{Radiative Modelling} \label{sec:level1}
In this section, we consider an analytic model of emissivity and absorptivity coefficients for the radiative transfer calculations. In what follows, we describe the prescription for radiative transfer \citep{2020ApJ...897..148G}. We  adopt the BH mass (M) for Sgr A* to be $4.3$ x $10^6 M_\odot$ and source distance (D) from the Earth is $8200 pc$ \cite{2019Sci...365..664D,2020A&A...636L...5G} for radiative transfer calculations.
Covariant form of the general relativistic radiative transfer is expressed as  
\begin{equation}
\frac{d\mathcal{I}}{d\tau_\nu} = -\mathcal{I} + \frac{{\eta}}{\chi},
\end{equation}
where $\mathcal{I}$ is the Lorentz-invariant intensity and is related to the specific intensity via $\mathcal{I} = I_\nu/\nu^3 = I_{\nu_0}/\nu_0^3$ where the subscript `0' denotes quantities in the local rest frame. $\tau_\nu$ is defined as the optical depth. $\chi$ and $\eta$ are the invariant absorption coefficient and emission coefficient at frequency $\nu$. Number density of the fluid is given by
\begin{equation}
    N = n_0 \exp{\left\{-0.5\left[\left(\frac{r}{10}\right)^2 + z^2\right]\right\}},
\end{equation}
where $z=h\cos\theta$. $n_0$ is the reference number density and $h$ is the vertical scale height.
We use the Keplerian angular momentum profile $\Omega = 1/r^{3/2}$ and the fluid four-velocity is thus given by 
\begin{equation}
    u_\mu = \bar{u}(-1, 0, 0, \Omega),
\end{equation}
with $\bar{u} = \sqrt{-(g^{tt} + g^{\phi\phi}\Omega^2 - 2g^{t\phi}\Omega)}$ such that $u_\mu u^\mu = -1$. The specific emissivity and absorption coefficients is given by
\begin{equation}
    j_\nu = \mathcal{C}N\left(\frac{v}{v_{obs}}\right)^{-\alpha}
        \end{equation}
\begin{equation}
    \alpha_\nu = \mathcal{A}\mathcal{C}N\left(\frac{v}{v_{obs}}\right)^{-(\alpha+\beta)},
\end{equation}
where $\mathcal{C}$ and $\mathcal{A}$ are constants controlling the effect of absorption and emission coefficients.
The Lorentz-invariant coefficients are then
\begin{align}
    \eta &= \frac{j_\nu}{\nu^2}, \\
    \chi &= \nu\alpha_\nu.
\end{align}
Table (\ref{table:model}) shows the values of different parameters for the accretion and the models of naked singularities.
\begin{table}
	\centering
	\caption{Model details}
	\label{table:model}
	\begin{tabular}{lllll} 
		\hline 
		\hline 
		\textbf{Parameter} & \textbf{Model-I} &\textbf{Model-II} &\textbf{Model-III} &\textbf{Sch. BH} \\
		\hline
		$M (M_\odot)$              & $4.3$ x $10^6$              &  $4.3$ x $10^6$               &   $4.3$ x $10^6$   &  $4.3$ x $10^6$\\
		$M_0$            & 0.70             &  0.63             &   -   &  -\\	
        $R_b$            & 2.857            &  3.175            &   -   &  -\\
        $\gamma$         & -                &  -                & 0.51  &  -\\
        $\mathcal{A}$    & 100              &  100              & 100    & 100\\
        $\mathcal{C}n_0$ (x$10^{-17}$) & 2.5 &  2.0 & $10^{-3}$ & 2.0\\
        $\alpha$         & -2               &  -2               & -2      & -2\\
        $\beta$          & 2.5              &  2.5              &  2.5    & 2.5\\
        $h$              & 10/3             &  10/3             &  10/3     & 10/3\\
	    \hline
	\end{tabular}
\end{table}
\begin{figure*} 
\centering
\includegraphics[scale=0.11]{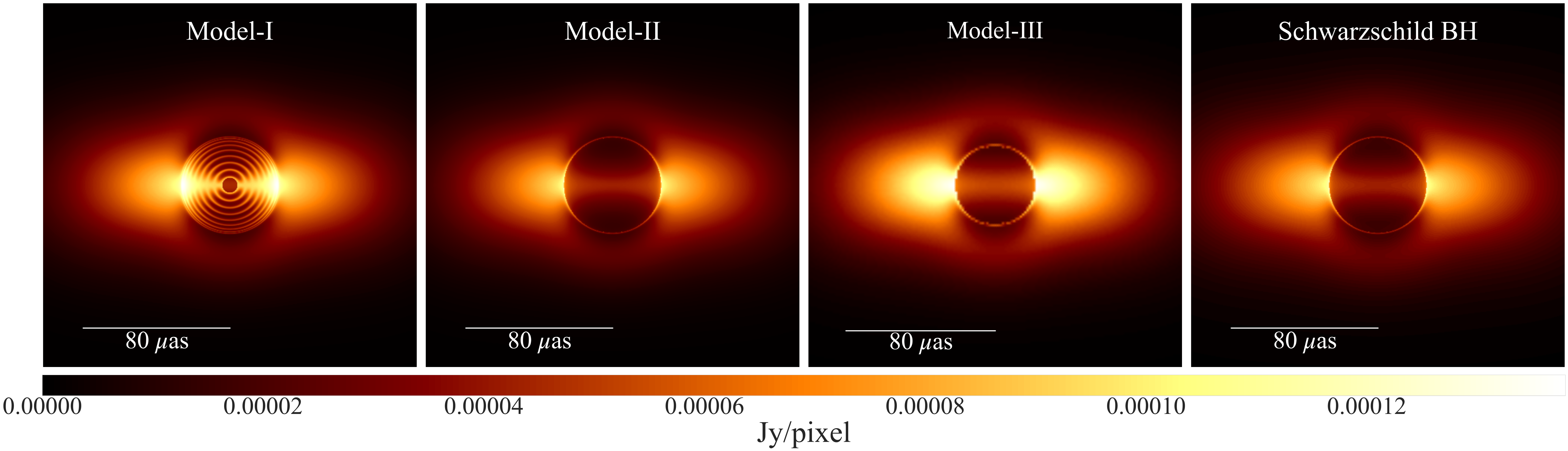}
\caption{Model images for Sgr A*, parameters details are given in the Table  (\ref{fig:model}).}
\label{fig:model}
\end{figure*}

\section{Ray-Tracing Formalism}
\label{sec:rtf}

In order to calculate the shadow image of a black hole or naked singularity, one must first
solve the geodesic equations in the background spacetime under
consideration. Here we describe the approach for ray-traced images of the models described in the previous section. We solve a system of 6 differential equations ($\dot{t}, \dot{r}, \dot{\theta}, \dot{\phi}, \dot{p_r}, \dot{p_\theta})$.

\subsection{Geodesic equations of motion}

For a given metric $g_{\alpha \beta}$, the Lagrangian can be written as
\begin{equation}
2\mathcal{L} = g_{\alpha \beta}\,\dot{x}^{\alpha}\dot{x}^{\beta}, 
\end{equation}
where an overdot denotes differentiation with respect to the affine
parameter, $\lambda$. 
From the Lagrangian, the covariant four-momenta of a geodesic may be written as:
\begin{equation}
p_\alpha = \frac{\partial \mathcal{L}}{\partial \dot{x}^\alpha}.
\end{equation}
Using the conservation of energy  and angular momentum, this can be broken down to write $p_t = -E$ and $p_\phi=L$, where, $E$ is the total energy of the particle and $L$ is the angular momentum in the direction of $\phi$. The final set of geodesic equations which will  be solved are then given by
\begin{align}
\dot{t} &= \frac{E}{g_{tt}} \\
\dot{r} &= \frac{p_r}{g_{rr}}\\
\dot{\theta} &=   \frac{p_\theta}{g_{\theta\theta}}\\
\dot{\phi} &= \frac{L}{g_{\phi\phi}} \\
\dot{p_r} &= \frac{1}{2g_{tt}}\left[ -{p_r}^2\left(\frac{g_{tt}}{g_{rr}}\right)'- (Q +  L^2)\left(\frac{g_{tt}}{g_{rr}}\right)' \right]  \\
\dot{p_\theta}  &= \frac{L^2}{g_{\theta\theta}}\frac{\cos{\theta}}{{\sin^3{\theta}}}
\end{align}
where $(')$ denotes differentiation with respect to $r$ and $Q$ is the Carter's constant which is the third constant of motion.

\subsection{Initial Conditions}
Here we discuss the initial conditions which are required to solve the system of an ordinary differential equation (ODE) described  in the previous section. In literature different authors have assumed different initial conditions based on the type of work and ODE's which  are needed to be solved. Here we use the formalism described in \cite{2016PhRvD..94h4025Y}. An observer is at some distance from the source. The observer is placed far away from the Black hole   ($r_{\mathrm{obs}} = 10^{3}\,M$), where the spacetime is assumed to be  flat (asymptotic flatness). The observer's position is specified in Boyer-Lindquist (oblate spheroidal)
coordinates as $(r_{\mathrm{obs}}, \theta_{\mathrm{obs}},  \phi_{\mathrm{obs}})$ where $\theta_{obs}$ and $\phi_{obs}$ are the inclination and azimuthal angles of the observer.

The transformation from Cartesian coordinates to Boyer-Lindquist coordinates is given by
\begin{align}
r^{2} &= \sigma+\sqrt{\sigma^{2}+a^{2}Z^{2}} \,, \label{CartBLr} \\
\cos \theta &= Z/r \,, \label{CartBLtheta} \\
\tan \phi &= Y/X \,, \label{CartBLphi}
\end{align}
where
\begin{align}
X &\equiv \mathcal{D}\cos\phi_{\mathrm{obs}}-x\sin\phi_{\mathrm{obs}} \,, \\
Y &\equiv \mathcal{D}\sin\phi_{\mathrm{obs}}+x\cos\phi_{\mathrm{obs}} \,, \\
Z &\equiv r_{\mathrm{obs}}\cos\theta_{\mathrm{obs}}+y\sin\theta_{\mathrm{obs}} \,,
\end{align}
and
\begin{align}
\sigma &\equiv \left(X^{2}+Y^{2}+Z^{2}-a^{2}\right)/2 \,, \\
\mathcal{D}
&\equiv \sin\theta_{\mathrm{obs}} \sqrt{r_{\mathrm{obs}}^{2}+a^{2}} -
y \cos\theta_{\mathrm{obs}}\,.
\end{align}
Finally, to obtain the ray’s velocity components in Boyer- Lindquist coordinates, we differentiate Eqs.(\ref{CartBLr})--(\ref{CartBLphi}), yielding:
\begin{align}
-\Sigma \ \dot{x}^{r} &= r \mathcal{R} \sin\theta \sin\theta_{\mathrm{obs}} \cos \Phi+\mathcal{R}^{2} \cos\theta \cos\theta_{\mathrm{obs}} \,, \label{rdot_initial} \\
-\Sigma \ \dot{x}^{\theta} &= \mathcal{R} \cos\theta \sin\theta_{\mathrm{obs}} \cos \Phi - r \sin\theta \cos \theta_{\mathrm{obs}} \,, \label{thetadot_initial}\\
\mathcal{R} \ \dot{x}^{\phi} &= \sin\theta_{\mathrm{obs}} \sin \Phi \ \!  \mathrm{cosec}\theta \,, \label{phidot_initial}
\end{align}
where
\begin{align}
\Sigma &\equiv r^{2} +a^{2}\cos^{2}\theta \,, \\
\mathcal{R} &\equiv \sqrt{r^{2}+a^{2}} \,, \\
\Phi &\equiv \phi-\phi_{\mathrm{obs}} \,.
\end{align}
Now, we have initial conditions for $(t, r, \theta, \phi, p_r, p_\theta)$. Using the above formalism, we generate the images for various models described in previous section and parameter values in Table~\ref{table:model}. Different columns in Figure~\ref{fig:model} depict these ray-traced models. Contrary to Model-II and Schwazschild BH, Model-I produces a very distinct image having multiple Einstein rings without casting a shadow, and a similar full-moon type image was produced previously by the authors in \cite{2019MNRAS.482...52S}.  Model-III on the other hand casts a relatively smaller shadow than Model-I and Schwzrscshild BH.

\begin{table}
	\centering
	\caption{Locations of Existing/Future Sites in the Event Horizon Telescope Array.}
	\label{table:sites}
	\begin{tabular}{lccr} 
		\hline
		\hline
		\textbf{EHT-I}    &    \textbf{Latitude}    &    \textbf{Longitude}    &    \textbf{SEFD.}$^{\dagger}$ \\
		\textbf{}    &    \textbf{($\deg$)}    &    \textbf{($\deg$)}    &    \textbf{(Jy)} \\
		\hline
		
		ALMA & -23.03 & -67.75 & 90    \\
        APEX & -23.01 & -67.76 & 3500  \\
        JCMT & 19.82 & -155.48 & 6000  \\
        LMT & 18.98 & -97.31 & 600   \\
        IRAM$^{\star}$    & 36.88  & -3.39   & 1400  \\
        SMA & 19.82 & -155.48 & 4900  \\
        SMT & 32.70 & -109.89 & 5000  \\
        SPT & -90.00 & 45.00 & 5000  \\
        \hline
        \hline
        \textbf{EHT-II, (Additional)}    &    \textbf{Latitude}    &    \textbf{Longitude}    &    \textbf{SEFD.}$^{\dagger}$ \\
        \textbf{}    &    \textbf{($\deg$)}    &    \textbf{($\deg$)}    &    \textbf{(Jy)}\\
        \hline
   
        GLT & 76.54 & -68.69 & 10000 \\
        KP & 31.96 & -111.61 & 10000 \\
        NOEMA & 44.63 & 5.91 & 700 \\
        \hline
        \hline
        \textbf{EHT-III, (Additional)}    &    \textbf{Latitude}    &    \textbf{Longitude}    &    \textbf{SEFD.}$^{\dagger}$ \\
        \textbf{}    &    \textbf{($\deg$)}    &    \textbf{($\deg$)}    &    \textbf{(Jy)} \\
        \hline 
        
        BAJA  & 30.87  & -115.46 & 10000 \\
        BOL   & -16.25 & -68.13  & 10000 \\
        CARMA & 37.1   & -118.14 & 10000 \\
        DRAK  & -29.3  & 29.27   & 10000 \\
        GAM   & 23.25  & 16.17   & 10000 \\
        HAY   & 42.43  & -71.49  & 2500  \\
        KAUAI & 21.79  & -159.51 & 10000 \\
        KEN   & -0.15  & 37.31   & 10000 \\
        PDB{}   & 44.44  & 5.91    & 1500  \\
        PIKES & 38.65  & -105.04 & 10000 \\
        VLT   & -24.48 & -70.4   & 10000 \\
		\hline
		\hline
	\end{tabular}
	
     $\star$ sites used for the observation only in EHT-I, III. \\
     $^{\dagger}$System Equivalent Flux Density

\end{table}

\begin{figure*}
\centering
\includegraphics[scale=0.245]{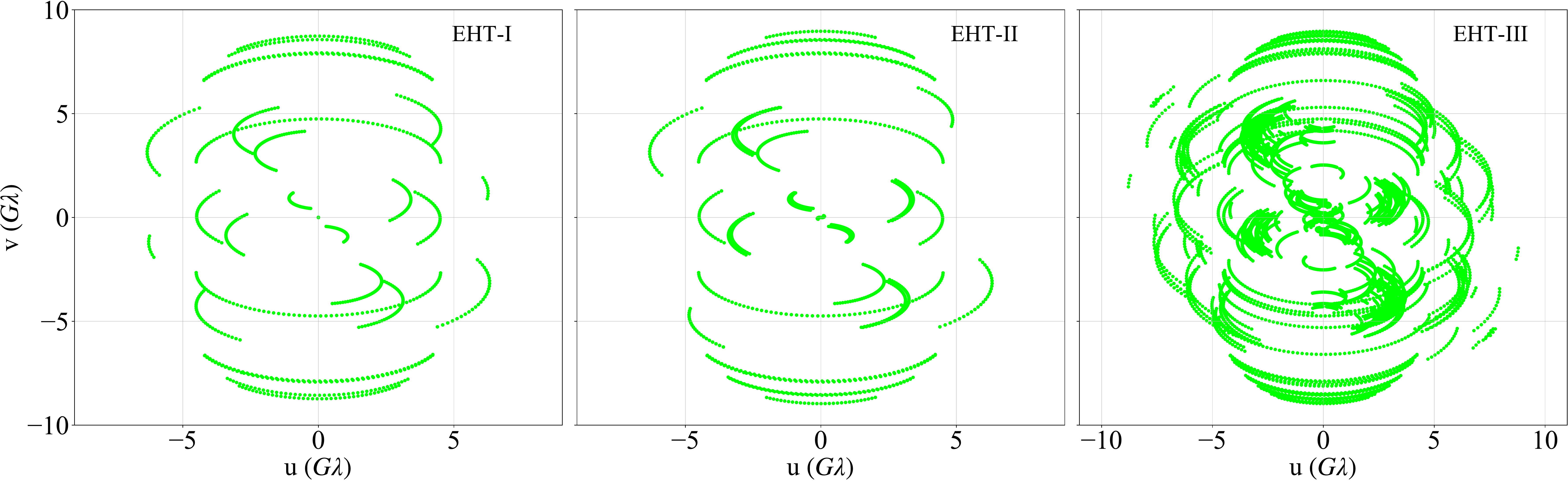}
\caption{Baseline coverage of Sgr A$^\star$ for the array configuration of EHT mentioned in Table~\ref{table:sites}.}
\label{fig:uv}
\end{figure*}


\begin{figure*}
\includegraphics[scale=0.12]{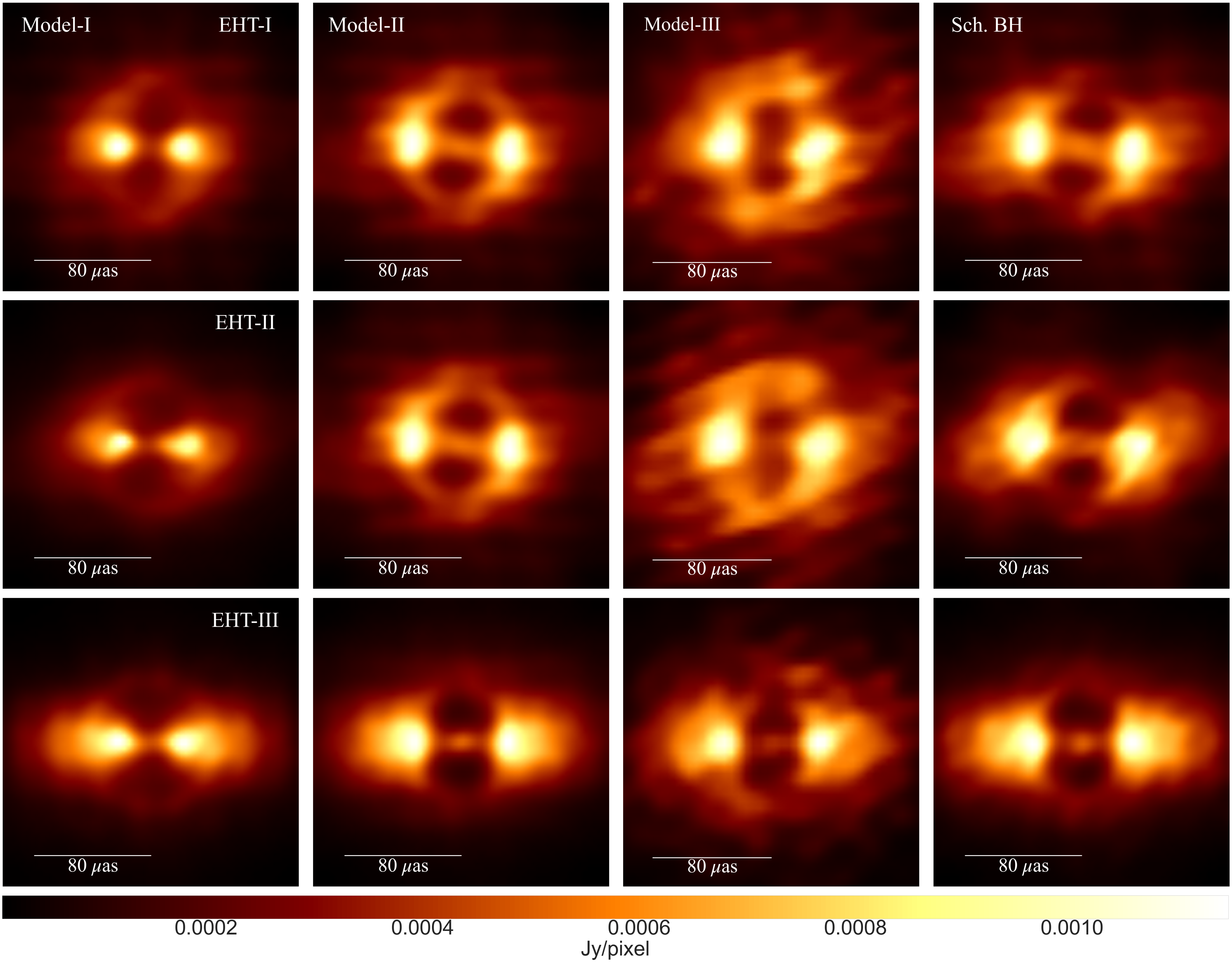}
\caption{Reconstructed images of Sgr A* for models described in section-I with different array configurations of EHT (top to bottom: EHT-I, EHT-II, EHT-III).}
\label{fig:rec-images}
 \end{figure*}
 

\section{EHT VLBI Observations}
\label{VLBI}
Continuing the analysis of the previous section, here we take the next step of visualizing these images with VLBI.
For this purpose, we utilize three different array configurations including current and future observations. We label them as EHT-I, EHT-II and EHT-III for 2017, 2021 and 2025 observations respectively as mentioned in Table ~(\ref{table:sites}). It should be noted that EHT-III is one of the possible configurations for future (2025) observations. Meanwhile, EHT-II has completed the observations for 2021 in early April. The images produced in the preceding section are infinite-resolution in the sense that they are produced by a pristine source that encounters no interstellar matter on its way to a perfect detector that can detect every photon that passes through it. In reality, interstellar scattering will occur, and the baselines will only cover a portion of the observing region.

We generate the synthetic radio images with $\texttt{ehtim}$ \cite{2018ApJ...857...23C} using the array configurations mentioned in Table~(\ref{table:sites}). The following parameters were used in the simulations: $\Delta\nu = 4$ GHz bandwidth, $t = 24$ hours, corresponding to a full day, at a central frequency of $230$ GHz. The total observing time is one of the most important parameters in the imaging process and this unusually long observation time allows us to present the ideal-case scenario. The (u, v) coverage grows larger as the observation time increases, allowing us to reconstruct a better image. We use the Maximum Entropy Method (MEM) to reconstruct static images from synthetic VLBI data.
Following this procedure, we perform the synthetic observations at the Galactic center. The visibility amplitudes are calculated by Fourier transforming the images and sampling them over the projected baselines of different arrays (EHT-I, EHT-II and EHT-III). To mimic realistic observations during the simulations, we include the effects of thermal noise and phase errors as well. As it can be seen in Fig. (\ref{fig:uv}), the (u,v) coverage is subsequently increasing as more antennas are added to the configuration. Images reconstructed for different models and arrays are presented in Fig. (\ref{fig:rec-images}). It should also be noted that the same imaging script was used for all the reconstructions used here. 
In the next sections, we do a qualitative and quantitative analysis of the results produced in this section

\subsection{Qualitative analysis}

Overall visual inspection of the images clearly shows the difference in the images of different models with different arrays. Starting from the left-hand side, as discussed in the previous section Model-I produces an interesting 'concentric rings type image. So the corresponding synthetic image for different arrays does not give a shadow-type structure. Due to the spherically symmetric nature of the model, the image in the third row for EHT-III shows a nearly circular structure. In fact, those multiple rings do not appear anymore in the reconstruction. This aberration is due to the resolution limit, making it very difficult to resolve them.

For Model-II, it casts a shadow and is clearly visible in all the arrays. Due to additional emissions in the line of sight, the shadow structure appears to be divided.  Although, it should be noted that it is a consequence of the radiative transfer modeling and raytracing that we see such type of emission along with the particular inclination. Clearly, there are differences between Model-I and Model-II in the image structure. Comparing Model-II with Schwarzschild BH, it can be seen that due to the same shadow size and similar emission regions, the difference in the images is not that significant. Hence, we can say that Model-II and Schwarzschild BH mimic each other, and one cannot tell the difference between them from the VLBI images.

For Model-III, the shadow size is smaller than the other models prescribed here which can be seen from the ray-traced images. So, in the EHT-I image, the shadow region is not as noticeable. The same may be said for EHT-II. The situation improves substantially with EHT-III, and the shadow region becomes more apparent. Model-III appears to be distinct from the Black Hole case. Even so, the change isn't all that substantial. This coincides with the previous case as well for Model-II. On a relative scale, Model-III is more distinct than Model-II.

\subsection{Quantitative Analysis}

Image-comparison measures, such as the Normalised Cross-correlation coefficient (NCC), structural dissimilarity index (DSSIM; Wang et al. 2004) etc., can be used to make a more quantitative evaluation of the degree of similarity among the various images under consideration. For this purpose here, we utilise NCC to compare the resultant images.
If two images are similar to each other, the maximum value NCC can have is 1, and the maximum dissimilarity will give NCC value to be around 0.
We incorporate this for three different arrays, namely EHT-I, EHT-II and EHT-III. NCC is then computed between the corresponding reconstructed Schwarzschild BH image with that of the other models as seen in Figure~\ref{fig:rec-images}. In that way, lower NCC values would suggest, that these  models can be differentiated from the Schwarzschild case. Given that, Model-I and Model-III produce distinct features (as explained in previous section), it is expected to deduce a similar result from NCC as well.
Table~\ref{table:comp} summaries these results. Model-I and Model-III give relatively lower values than Model-II, suggesting that the images are distinguishable.
Model-I having no shadow and Model-III having a smaller shadow size  as shown in  Figure~\ref{fig:model} are responsible for various structures in the reconstructed images resulting in relatively lower values of NCC than the Model-II which has the same shadow size as Schwarzschild BH.

\begin{table}
	\centering
	\caption{Normalised Cross-Correlation coefficient for the comparison between the reconstructed of various models with the Schwarzschld BH.}
	\vspace{0.35cm}
	\label{table:comp}
	\begin{tabular}{llll} 
		\hline 
		\hline 
		\textbf{Sch. BH}    &   \textbf{Model-I}    &    \textbf{Model-II}     &     \textbf{Model-III}\\
		\hline
		\textbf{EHT-I}    & 0.88 &  0.96 & 0.74 \\	
        \textbf{EHT-II}   & 0.90 &  0.98 & 0.73 \\
        \textbf{EHT-III}  & 0.92 &  0.99 & 0.75 \\
        \hline
	    \hline
	\end{tabular}
\end{table}

\section{Discussions and Conclusions} 
\label{result}
Here, we have modeled Sgr A* as the JMN-1 or JNW naked singularity with the possibility of having either a shadow-like structure or a pristine source (concentric ring-like structure) emitting from all the regions. We use the ray-tracing algorithm to solve the geodesic equations and then solve the general relativistic radiative transfer equation. We use an analytic model for emission and absorption coefficients for generating emission maps. Afterward, we produce synthetic observations for the prescribed models of the source considering three different array configurations namely EHT-I, EHT-II and EHT-III and reconstruct the images.

Now, from the overall analysis, it is clearly seen that the naked singularity models are distinguishable from a Black hole with VLBI imaging. Both qualitative and quantitative analyses of the images provides us with a rough estimate of the differences between the models. Model-I and Model-II exhibit different accretion structures and imaging artifacts, which can be seen in the image reconstructions. Meanwhile, Model-II remains to be a Black hole mimicker and is almost similar to a Black hole in terms of size and emission region. The NCC quantifies this inference, and has values near to $1$ for only Model-II. However, for other models, the values are relatively smaller differentiating the models quantitatively. The NCC can be used to explore these potential differences to some extent.

As we have seen that the synthetic VLBI images for the EHT-I array for Model-II, Model-III and Schwarzschild BH give shadow-like features but on the other hand Model-I does not give any hint towards that due to the fact that it does not produce shadow theoretically. As we move on to increase the number of arrays in the configuration, the Fourier domain fills up significantly, generating a clear image. So, when we compare the models on the basis of increased arrays, the models retain their original state in the sense of shadow. 

The above analysis concludes that when Sgr A* is observed on these baselines and an image is reconstructed, if the image showcase a shadow-like structure, one cannot distinguish between a Black hole and a naked singularity. Moreover, if Sgr A* does not contain a BH but a naked singularity (Model-I) then one can expect an image without shadow as seen in Fig ~[\ref{fig:rec-images}]. No matter how much the baseline is increased or the number of arrays are added there will not be any image having a shadow feature. Indeed, the absence of the shadow in the JMN-1 spacetime introduces important and fundamental differences in the flow dynamics and images. It can act as a possible observable to test the presence of a naked singularity. This can thus allow us to probe the nature of Sgr A* with shadows and VLBI imaging. 

Although the analysis presented here is based on the images, certainly there are a few caveats involved which require certain attention. Firstly, the radiative modeling that we have used for the accretion region is a simple analytical model. The number density that we have used can be thought to be analogous to the electron number density in RIAF modeling. RIAF spatial distributions of the electron temperature $T_e$ and thermal electron density $n_e$ are usually described by a hybrid combination \cite{2018ApJ...863..148P}. Different eDF's (electron distribution function) are also incorporated in the emission and absorption coefficients to model the emission regions more accurately \cite{2016ApJ...822...34P}. The radiative cooling timeframe for RIAF is significantly longer than the accretion timescale for a cold, geometrically thin, Keplerian revolving disk, resulting in a hot, geometrically thick flow with a sub-Keplerian rotation (see for eg. \cite{1994ApJ...428L..13N} \& \cite{1997ApJ...476...49N}). Introducing different flow-dynamics can produce some interesting and distinguishable features in the resulting images \cite{2016ApJ...831....4P} as well. To conclude, these following caveats would be interesting topics to follow up the current work in the future. Additionally, they would provide us with a more robust way to probe the nature of Sgr A* for a number of observable.

Upcoming results from the 2017 and 2021 observational campaigns, may provide more insight into the argument. If the shadow is observed in the image of Sgr A*, we cannot conclusively say that it contains a Supermassive Black hole, as we have seen from the ray-traced and synthetic images of naked singularity models and a Black hole. Moreover, if the shadow is absent in the image, it would be a hint towards the presence of a naked singularity, but not a Black hole. Concerning the general relativity (and not the other modified gravities), these naked singularity solutions present a good alternative to the conventionally used Black hole.

\section*{Data Availability}
Data underlying related to the results of simulations in this article will be shared on reasonable request to the corresponding author.

\section*{Acknowledgements}
The authors would like to thank Yosuke Mizuno for helpful and valuable comments regarding the work.

\bibliographystyle{mnras} 

\bsp	
\label{lastpage}
\end{document}